\documentclass{aastex63}

\usepackage{amsmath}
\usepackage{amssymb}
\usepackage{graphicx}
\begin{document}

\title{Current-sheet Oscillations Caused by Kelvin-Helmholtz Instability at the Loop Top of Solar Flares}

\correspondingauthor{Xin Cheng}
\email{xincheng@nju.edu.cn}

\author[0000-0001-9863-5917]{Yulei Wang}
\affiliation{School of Astronomy and Space Science, Nanjing University, Nanjing 210023, People's Republic of China}
\affiliation{Key Laboratory for Modern Astronomy and Astrophysics (Nanjing University), Ministry of Education, Nanjing 210023, People's Republic of China}

\author[0000-0003-2837-7136]{Xin Cheng}
\affiliation{School of Astronomy and Space Science, Nanjing University, Nanjing 210023, People's Republic of China}
\affiliation{Key Laboratory for Modern Astronomy and Astrophysics (Nanjing University), Ministry of Education, Nanjing 210023, People's Republic of China}
\affiliation{Max Planck Institute for Solar System Research, Gottingen, D-37077, Germany}

\author{Zining Ren}
\affiliation{School of Astronomy and Space Science, Nanjing University, Nanjing 210023, People's Republic of China}
\affiliation{Key Laboratory for Modern Astronomy and Astrophysics (Nanjing University), Ministry of Education, Nanjing 210023, People's Republic of China}

\author[0000-0002-4978-4972]{Mingde Ding}
\affiliation{School of Astronomy and Space Science, Nanjing University, Nanjing 210023, People's Republic of China}
\affiliation{Key Laboratory for Modern Astronomy and Astrophysics (Nanjing University), Ministry of Education, Nanjing 210023, People's Republic of China}

\begin{abstract}
Current sheets (CSs), long stretching structures of magnetic reconnection above solar flare loops, are usually observed to oscillate, their origins, however, are still puzzled at present. 
Based on a high-resolution 2.5-dimensional MHD simulation of magnetic reconnection, we explore the formation mechanism of the CS oscillations. 
We find that large-amplitude transverse waves are excited by the Kelvin-Helmholtz instability (KHI) at the highly turbulent cusp-shaped region.
The perturbations propagate upward along the CS with a phase speed close to local Alfv\'{e}n speed thus resulting in the CS oscillations we observe.
Though the perturbations damp after propagating for a long distance, the CS oscillations are still detectable.
In terms of detected CS oscillations, with a combination of differential emission measure technique, we propose a new method for measuring the magnetic field strength of the CSs and its distribution in height.
\end{abstract}


\section{Introduction} \label{sec:intro}

Solar flares are one of the most energetic phenomena in the solar atmosphere and often appear as a sudden emission enhancement over the whole electromagnetic spectrum.
In the past decades, their energy release mechanism, spatial structures, and dynamical properties have been widely studied. 
The well-known standard flare model summarizes several key features of flares, including two parallel bright ribbons, cusp-shaped loop-top structure, elongated current sheet (CS), and the erupting magnetized plasmoid \citep[CSHKP;][]{CarmichaelH_1964_729,SturrockPA_1966_968,HirayamaT_1974_969,KoppRA_1976_727,ShibataK_1995_814,LinJun_2000_709}.
These features are all closely related to the fundamental energy release process, i.e., magnetic reconnection.

Above the flare loops, the thin and long stretching or ray-like high-temperature structures, typically of 10-20 million Kelvins, are suggestive of model-predicted CSs \citep[see][]{Ciaravella2002,Webb2003,Ko2003,Liu2010a,Patsourakos2011}.
For a few events, it is observed that the thin and long CSs even extend to the high corona for serval solar radii and connect to the erupting coronal mass ejections \citep[e.g.,][]{ChengXin_2018_341}.
Because of the large Lundquist number of the corona, plenty of small-scale plasmoid structures are generated by tearing mode instability and then propagate upward and downward along the CSs \citep[e.g.,][]{ShenChengCai_2011_619,MeiZhiXing_2012_614,LinJun_2015_723,LeeJaeOk_2021_850}.

\begin{figure*}[h]
\gridline{
    \fig{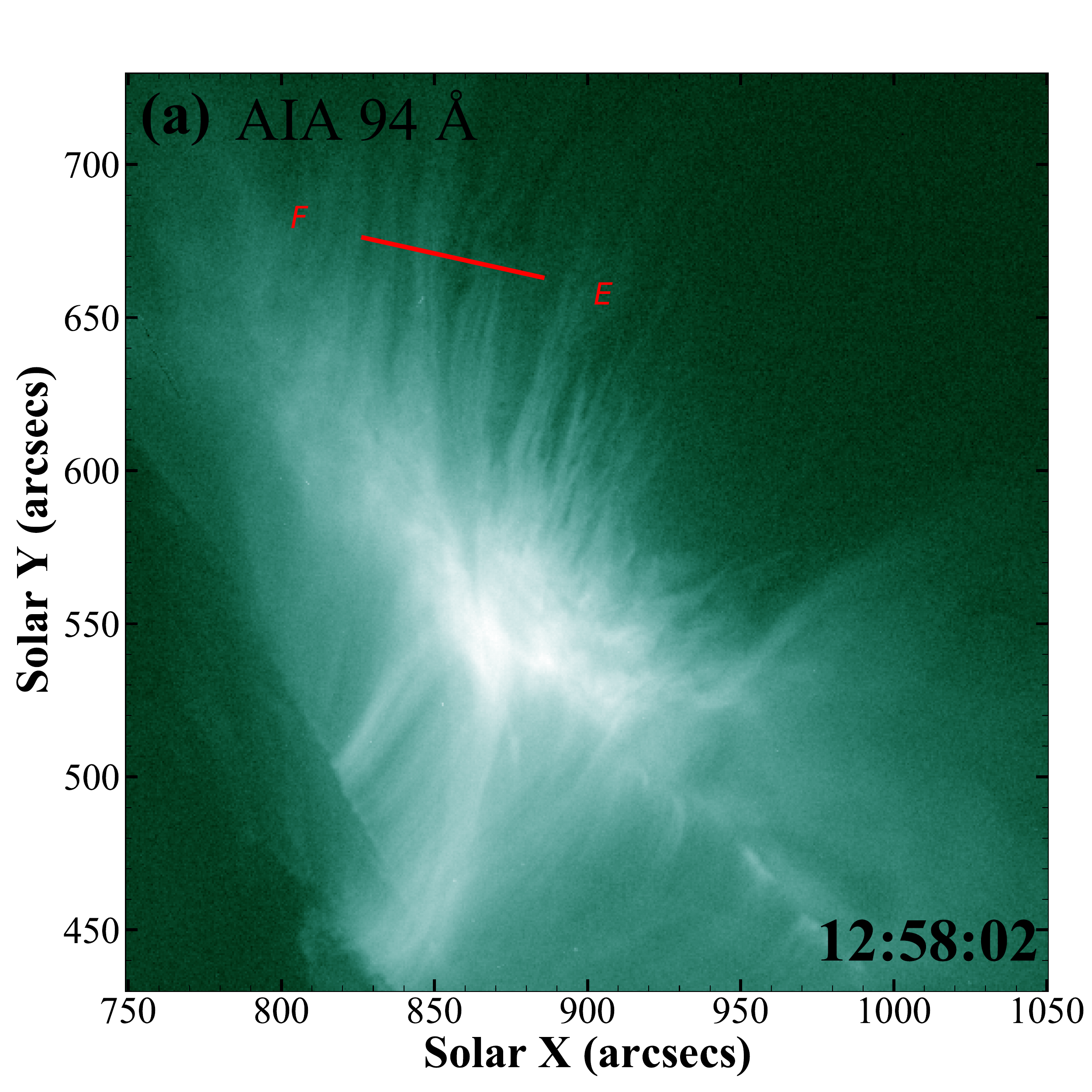}{0.39\textwidth}{}
    \fig{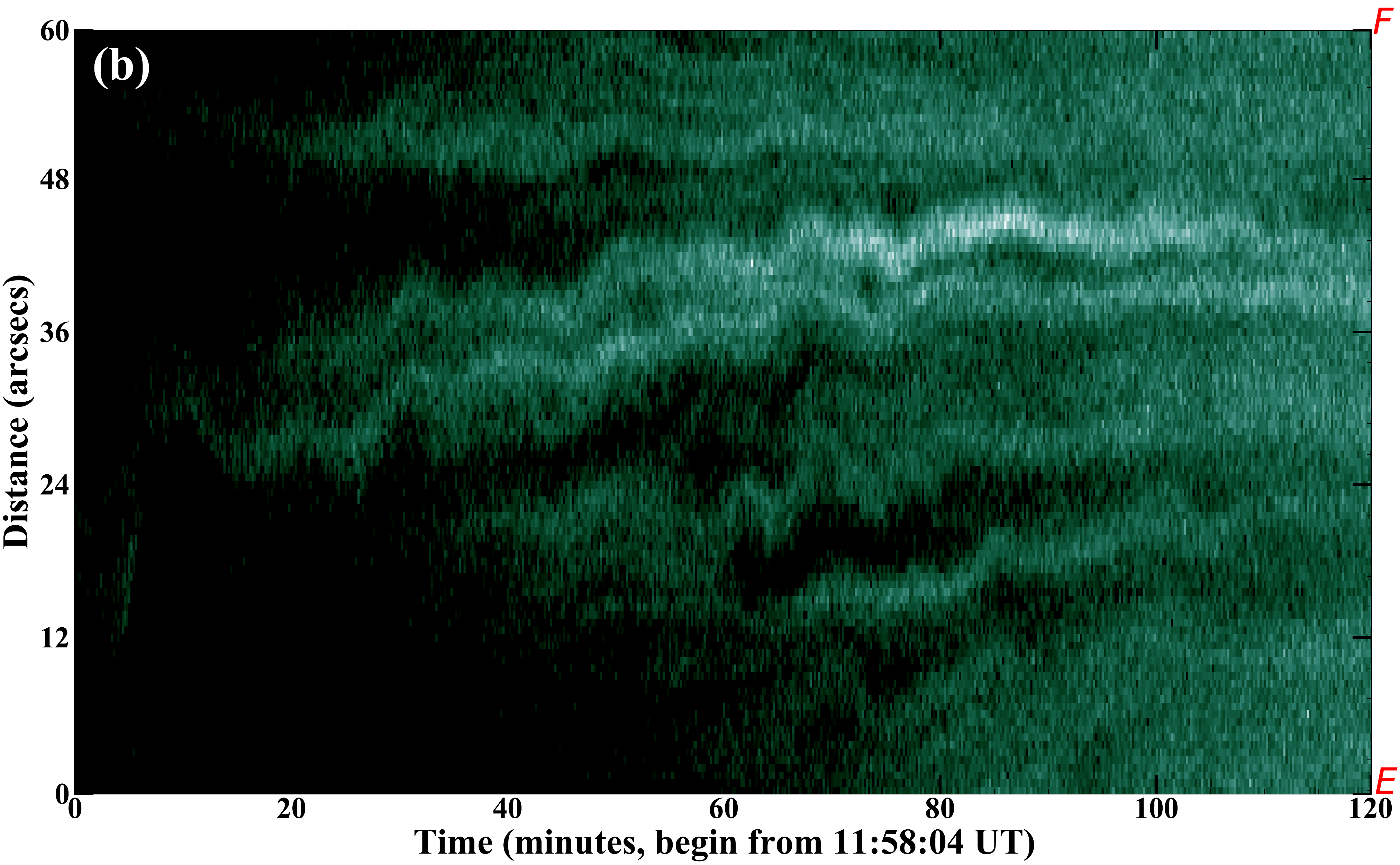}{0.61\textwidth}{}
}
\caption{A loop-top CS oscillation event.
(a) AIA 94\,\AA\ image showing the ray-like structures above flare loops recorded at 2011-10-22 12:58:02 UT suggestive of the elongated CSs \citep[see also][]{Webb2003,LiLeping_2016_844}.
(b) Slice-time plot showing the oscillations of the ray-like CSs. The location of the slice EF is indicated in panel (a).
An animation of this figure is available online.
The video, with a duration of $15\,\mathrm{s}$, shows the evolution of the ray-like structures in the 94\,\AA\ passband from 11:58:04 to 13:57:04 on 2011-10-22.
\label{fig0}}
\end{figure*}

Interestingly, the CSs usually also present a transverse oscillation.
\cite{Chen2010} reported propagating large-scale waves along a helmet streamer CS with a period of one hour and wave-length of several solar radii.
Similar waves with a shorter period of $30\,\mathrm{min}$ were also captured by \cite{LingAG_2014_842}.
\cite{SamantaTanmoy_2019_752} analyzed a similar but more slowly propagating wave ($\sim 19\,\mathrm{km/s}$) along the ray-like CSs and proposed that they might be vortex shedding in the corona.
\cite{Verwichte2005} found that the supra-arcade downflows (SADs), probably corresponding to the reconnection outflows, as well as the bright CS-like supra-arcade fans (SAFs), also oscillated but the periods were usually much shorter, mostly $<5\,\mathrm{min}$.
The first evidence for the CS oscillations was reported by \cite{LiLeping_2016_844}, who found that it had a period of $\sim 11\,\mathrm{min}$ and propagated away from the sun with a speed of $\sim 200\,\mathrm{km/s}$.
Figure \ref{fig0} exhibits a similar CS oscillation event captured by the Atmospheric Imaging Assembly (AIA) on board the \emph{Solar Dynamic Observatory} (SDO) \citep{PesnellWDean_2012_970,LemenJamesR_2012_951} on 2011 Oct. 22. 
One can clearly see that the ray-like structures above the flare loops, supposed to be the CSs, sway in the plane of the sky.

The CS oscillations might be related to the fine processes at the cusp-shaped flare loop region where turbulent flows and plenty of shocks take place \citep[see][]{McKenzieDE_2013_745,TakasaoShinsuke_2015_717,ShenChengcai_2018_705,CaiQiangwei_2019_872,YeJing_2020_653}.
\cite{TakahashiTakuya_2017_824} reported that, under high-Lundquist-number conditions, the quasi-periodic oscillations were driven by the horizontal motions of termination shocks (TSs) with oblique fronts at the flare loop top and the buffer region at the CME bottom.
Recently, \cite{XieXiaoyan_2021_977} performed a similar simulation and found that the oscillations are caused by Rayleigh-Taylor instability (RTI) in the buffer region with the break of symmetry and that the oscillations propagated towards the sun and caused prominent displacements of the CS structure.

In this letter, we further explore the origins of the CS oscillations based on a high-resolution 2.5-dimensional magnetic reconnection model. 
We found that the Kelvin-Helmholtz instability (KHI) of the TS tail flow initializes the development of asymmetry flows and thus gives rise to the CS oscillations.
Furthermore, we analyze the propagation properties of perturbations causing the CS oscillations and confirm the method we propose for estimating the magnetic field strength of the CS and its distribution in height.
Section \ref{sec:method} briefly introduces our numerical model. 
The main results are presented in Section \ref{sec:results}, which are followed by a summary and discussion. 

\section{Numerical Model\label{sec:method}}

Our numerical model is based on the resistive MHD equations including gravity, anisotropic thermal condition, radiation cooling, and background heating:
\begin{eqnarray}
\frac{\partial\rho}{\partial t}+\nabla\cdot\left(\rho{\bf u}\right) & = & 0\,,\nonumber \\
\frac{\partial\left(\rho{\bf u}\right)}{\partial t}+\nabla\cdot\left(\rho{\bf u}{\bf u}-{\bf BB}+P^{*}{\bf I}\right) & = & \rho\bf{g}\,,\nonumber \\
\frac{\partial e}{\partial t}+\nabla\cdot\left[\left(e+P^{*}\right){\bf u}-{\bf B}\left(\bf{B}\cdot{\bf u}\right)\right] & = & \rho{\bf g}\cdot {\bf u}+\nabla\cdot\left(\kappa_{\parallel}\hat{\bf{b}}\hat{\bf{b}}\cdot\nabla T\right)-n_in_e\Lambda\left(T\right)+H\,,\label{eq:MHD}\\
\frac{\partial{\bf B}}{\partial t}-\nabla\times\left(\bf{u}\times\bf{B}\right) & = & -\nabla\times\left(\eta{\bf J}\right)\,,\nonumber \\
{\bf J} & = & \nabla\times{\bf B}\,,\nonumber
\end{eqnarray}
where, $P^*=p+B^2/2$, $e=p/\left(\gamma-1\right)+\rho u^2/2+B^2/2$, $\gamma=5/3$, $n_e$ and $n_i$ are respectively the number density of electrons and ions, and other variables are denoted by standard notations.
The conductivity parallel with magnetic field is determined by $\kappa_\parallel=\kappa_0T^{2.5}$, where $\kappa_0=6.67\times10^5\,\mathrm{erg\cdot s^{-1}\,cm^{-1}\,K^{-3.5}}$.
The gravity acceleration is calculated by ${\bf g}=-g\hat{\bf{e}}_y$, where $g=g_0/\left(1+y/R_{\sun}\right)^2$, $R_{\sun}$ denotes the solar radii, and $g_0=2.7390\times 10^4\,\mathrm{cm/s^2}$.
We use a widely-used optically thin radiation cooling function $\Lambda\left(T\right)$ \citep[see][]{KlimchukJA_2008_961,YeJing_2020_653,Shen2022}.
The background heating, $H=n_in_e\Lambda\left(T_\mathrm{cor}\right)$, is supposed to maintain the initial energy balance in corona and also keep balancing the cooling effect in the bottom chromosphere region ($y<0.2$), where $T_\mathrm{cor}$ is the initial coronal temperature.
In this paper, all variables are normalized according to constant units identical to \cite{WYL_2021_964}.
Physical units of main variables are listed in Table \ref{tab:Units}.

\begin{table}[h]
    \centering
    \begin{tabular}{lll}
    \hline
    \textbf{Units}        & \textbf{Symbols} & \textbf{Values} \tabularnewline
    \hline
    Space                 & $L_0$            & $50\,\mathrm{Mm}$\tabularnewline
    Mass density          & $\rho_0$          & $1.67\times10^{-14}\,\mathrm{g/cm^3}$\tabularnewline
    Magnetic strength     & $B_0$            & $20\,\mathrm{G}$\tabularnewline
    Time                  & $t_0$            & $114.61\,\mathrm{s}$\tabularnewline
    Velocity              & $u_0$            & $436\,\mathrm{km/s}$\tabularnewline
    Temperature           & $T_0$            & $11.52\,\mathrm{MK}$\tabularnewline
    \hline
    \end{tabular}
    \caption{Physical units of main variables.\label{tab:Units}}
\end{table}

The initial temperature of the gravitationally stratified atmosphere is set as 
\begin{equation}
    T\left(y\right)=\frac{T_\mathrm{cor}-T_\mathrm{chr}}{2}\mathrm{tanh}\left(\frac{y-h_\mathrm{chr}}{w_\mathrm{tr}}\right)+\frac{T_\mathrm{cor}+T_\mathrm{chr}}{2}\,,\label{eq:T}
\end{equation}
where, $T_\mathrm{cor}=0.1$, $T_\mathrm{chr}=0.002$, $h_\mathrm{chr}=0.12$, and $w_\mathrm{tr}=0.02$.
The initial pressure profile, calculated by $p\left(y\right)=p_0\mathrm{exp}\left(-\int_0^y g/T\mathrm{d}y'\right)$, is set to balance gravity \citep[see also][]{YeJing_2020_653}, where $p_0=p_\mathrm{ref}\mathrm{exp}\left(\int_0^{y_\mathrm{ref}}g/T\mathrm{d}y'\right)$ and $p_\mathrm{ref}=0.08$ is the pressure at $y_r=0.22$. 
The initial force-free magnetic field, identical to \cite{WYL_2021_964}, forms a vertical CS resembling that in the CSHKP model.
To trigger the reconnection, we use a localized anomalous resistivity near $y=0.5$ and $x=0$, which is also identical to Wang et al. (2021).
The initial velocity is set to zero everywhere.
The initial static equilibrium of the atmosphere is exactly satisfied in the corona region (y>0.2), but is slightly perturbed in the transition region (0.1<y<0.2) where the thermal conduction term causes a localized downward energy flow due to the temperature gradient.
However, this initial perturbation in the transition region is ignorable compared with the dominating reconnection process (see the animation of Figure 2).
The boundary conditions are arranged as follows.
The left ($x=-0.5$) and right ($x=0.5$) are free boundaries, the top ($y=2$) is no-inflow boundary, and the bottom ($y=0$) is symmetric boundary.

The above system is simulated with the \textsf{Athena++} code \citep{StoneJamesM_2020_941}.
We use the HLLD Riemann solver \citep{MiyoshiTakahiro_2005_876}, the 2-order piecewise linear method (PLM), and the 2-order van Leer predictor-corrector scheme to solve the conservation part of Eq.\,\ref{eq:MHD}. 
The resistivity, thermal conduction, gravity, radiation cooling, and heating terms are calculated by the explicit operator splitting method.
The 2-order RKL2 super-time-stepping algorithm is adopted to reduce computational costs \citep{MeyerChadD_2014_939}.
Uniform Cartesian grids are used on both directions, namely, 1920 and 3840 grids on $x$ and $y$ directions, respectively. The pixel scale is $\Delta L=\Delta x=\Delta y=26\,\mathrm{km}$.
The maximum simulation time in our simulation is $t_{max}=12$, which corresponds to $23$ minutes in physical time.

\section{Results\label{sec:results}}

\subsection{Overview\label{subsec:overview}}

\begin{figure*}[h]
\gridline{\fig{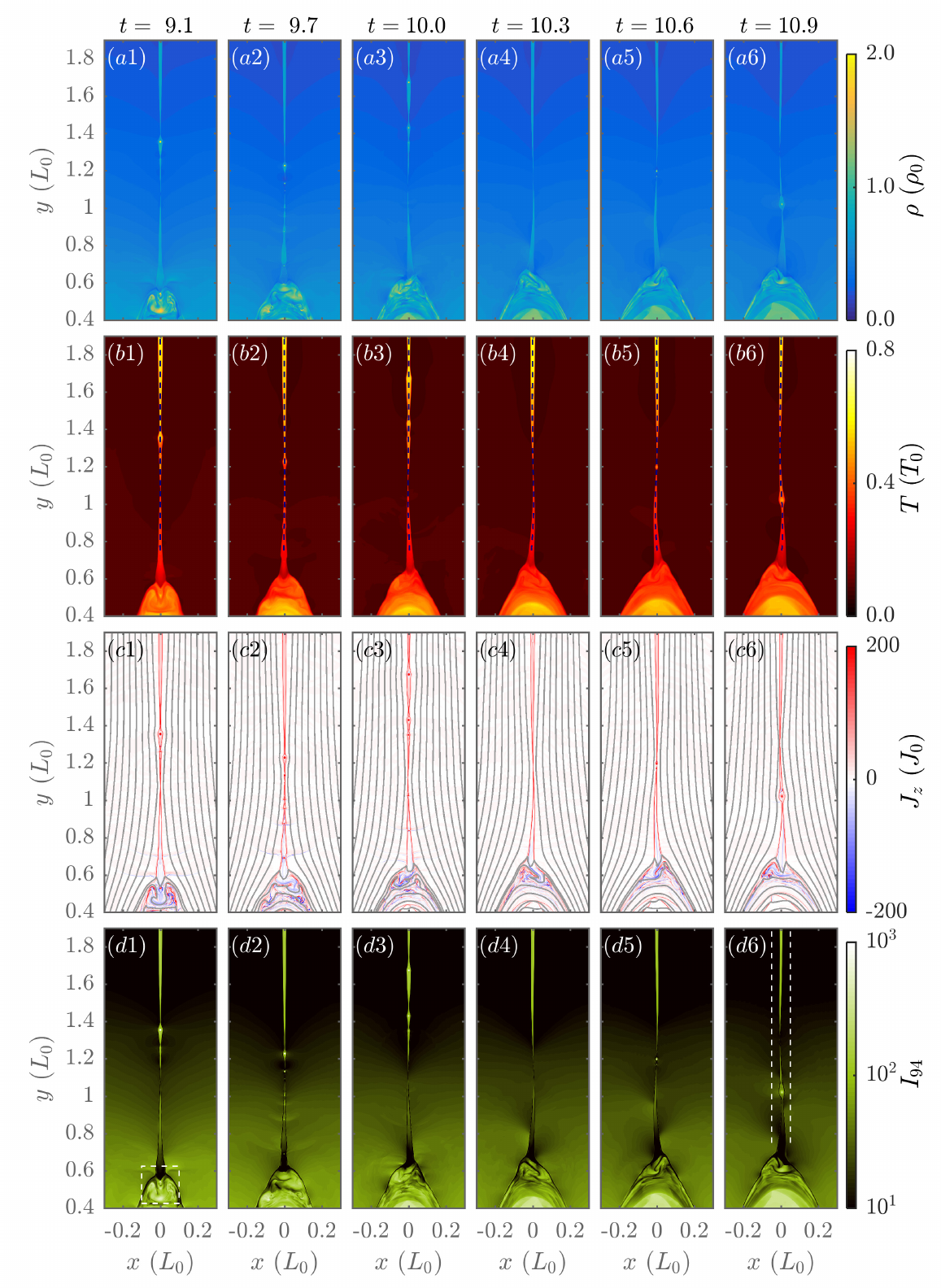}{0.8\textwidth}{}}
\caption{Distributions of density $\rho$, temperature $T$, current density $J_{z}$, and synthetic AIA 94\,\AA\ images at six moments.
The region is selected to emphasize the loop top region and the CS.
The blue dashed curves in the temperature profile (2nd row) are the middle of the CS.
The grey curves in panels (c1)-(c6) exhibit the magnetic field lines.
The dashed box in panel (d1) marks the region as shown in Figure \ref{fig2}.
The dashed lines in panel (d6) show the CS outer boundary used in Figure \ref{fig4}. 
The unit of the intensity for the 94\,\AA\ images is $\mathrm{DN\,pixel^{-1}s^{-1}}$.
An animation of this figure is available online.
The video, with a duration of $24\,\mathrm{s}$, shows the evolutions of $\rho$, $T$, $J_z$, and synthetic AIA 94\,\AA\ image in $t\in\left[0,12\right]$.
\label{fig1}}
\end{figure*}

The magnetic reconnection is initiated at a small region that has an anomalous resistivity.
At $t=5$, a flare loop system and an erupting plasmoid above the principal X-point appear \citep[see also][]{YokoyamaTakaaki_2001_731,TakasaoShinsuke_2015_717}.
Under the small background resistivity, plasmoid instability dominates the reconnection and the CS is fragmented into various magnetic islands of different scales \citep[see also][]{BhattacharjeeA_2009_682,ShenChengCai_2011_619,MeiZhiXing_2012_614,NiLei_2016_832,YeJing_2019_706,KongXiangLiang_2020_719,ZhaoXiaozhou_2020_679}.
The reconnection also shows a quasi-periodic characteristic.
At $t=7.5$, a relatively large magnetic island forms in the CS, it then moves downward and collides with the flare loop top (see the animation of Figure \ref{fig1}). 
As the island annihilates there gradually, the kinetic and magnetic energies are further released \citep[see][]{WYL_2021_964}. 
At the same time, the cusp-shaped loop top becomes highly turbulent.
Following this big island, several relatively small islands are generated and also enter the loop top region.
After $t=9$, large-size islands are rarely generated and the oscillations of CS gradually appear.  
At $t=9.1$, under the TS, the $y$-asymmetry of the flows starts to grow (see the 1st column of Figure \ref{fig1}).
The tip of the cusp-shaped loop top show prominent oscillation after $t=9.7$ (see Figure \ref{fig1}).
The CS bottom sways back and forth, and the transverse waves are excited and propagate upward at the same time.
To compare with the observations, following the method used by \cite{XieXiaoyan_2019_980} and \cite{YeJing_2020_653}, we synthesize the AIA 94\,\AA\ images by $I_{94}=\int n_e^2f\left(T\right)\mathrm{d}l=n_e^2f\left(T\right)L_{\mathrm{los}}$, where $f\left(T\right)$ is the temperature response function, and $L_{\mathrm{los}}=10^9\,\mathrm{cm}$ is the scale of the line-of-sight on $z$-direction. 
The wave-like swinging of the CS can be clearly observed in the synthesized AIA 94\,\AA\ images (See Figures \ref{fig1}(d1)-(d6)).

\subsection{Initialization of CS Oscillations\label{subsec:KHI}}

\begin{figure*}[h]
\gridline{\fig{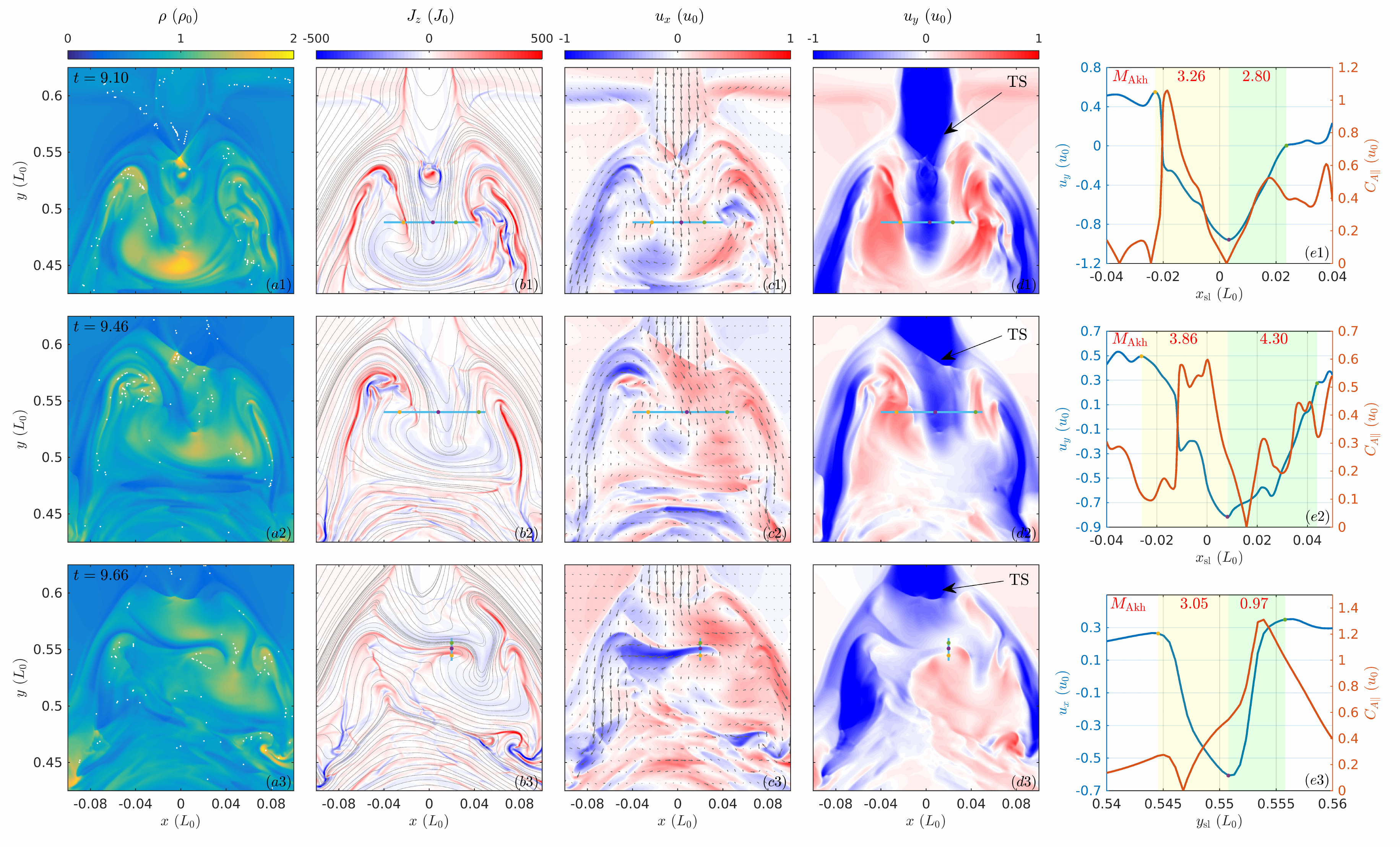}{1\textwidth}{}}
\caption{(a1)-(d1) Distributions of $\rho$, $J_z$, $u_x$, and $u_y$, where the blue line segments are approximately perpendicular to the shear layers.
(e1) The sheared velocity (blue curves) and $C_{A\parallel}$ (orange curve) along the line segments.
The positions of local velocity extreme values are marked by the orange, purple, and green dots, which are also used to define the ranges of shear layers as highlighted by the yellow and green shades in panel (e1).
The values of $M_{\mathrm{Akh}}$ are shown in red at the top of the panel.
The grey curves in panel (b1) denote the magnetic field lines, while the arrows in panel (c1) exhibit the velocity field. 
The white dotes in panel (a1) mark the positions of $\lambda_c<0.1$, namely, where the RTI can potentially be triggered.
(a2)-(e2) and (a3)-(e3) are the same as (a1)-(e1) but for different moments.
An animation of this figure is available online. 
The video, with a duration of $15\,\mathrm{s}$, shows the evolutions of $\rho$, $J_z$, $u_x$, and $u_y$ at the loop-top region in $t\in\left[9,10.5\right]$.
The positions of $\lambda_c<0.1$ are also marked in the movie of $\rho$.
\label{fig2}}
\end{figure*}

Figure \ref{fig2} exhibits the initialization phase of the CS swing at the cusp-shaped loop top.
Before $t=9.1$, the loop top is collided consecutively by bullet-like magnetic islands.
Though many small-scale vortexes emerge, the loop top is approximately symmetric in the $y$-direction and no global horizontal displacement is observed (see the animation of Figure \ref{fig2}).
At $t=9.1$, under the TS front (the downstream region), the downflow speed is still strong (see Figure \ref{fig2}(d1) and (e1)).
Once the tail flow is blocked by the relatively stable flare loops, two backflows form on both sides (e.g., Figure \ref{fig2}(c1) and (d1), also see \cite{TakasaoShinsuke_2016_716}).
Consequently, two shear layers form below the TS, where the KHI is triggered.

Theoretically, if the magnetic field is strong enough, the magnetic tension along the shear layer can stabilize the perturbations and thus suppress the KHI \citep{JonesTW_1997_978}.   
The key criterion of the KHI in magnetized plasmas is the local Alfv\'{e}nic Mach number of the velocity transition in shear layers \citep[e.g.,][]{RyuDongsu_2000_946}, which is defined as $M_{\mathrm{Akh}}=U_0/C_{A\parallel}$, where $U_0$ is the shear velocity, $C_{A\parallel}=B_\parallel/\sqrt{\rho}$ is the projected Alfv\'{e}n speed, and $B_\parallel$ is the magnetic strength along the shear layer.
To be specific, the KHI is stabilized by the magnetic tension for $M_{Akh}<2$, and if $M_{Akh}>4$, the magnetic field is too weak and the evolution of the KHI is almost fully hydrodynamical \citep[see][]{RyuDongsu_2000_946}. 

To estimate $M_{\mathrm{Akh}}$ of two shear layers, we set a slit perpendicular to the shear layers (see Figures \ref{fig2}(b1)-(d1)) and extract the profiles of the parallel velocity (here $u_y$) and $C_{A\parallel}$ along the slit (Figure \ref{fig2}(e1)).  
$U_0$ is the difference between the speeds of sheared flows (e.g., the velocity difference between the orange and purple dots in Figure \ref{fig2}(e1)).
Because $C_{A\parallel}$ varies along the slit, the average value across the shear layers is used when calculating $M_\mathrm{Akh}$.
As shown by Figure \ref{fig2}(e1), the values of $M_{\mathrm{Akh}}$ on the two shear layers (left and right) are $3.26$ and $2.80$, respectively, which means that the two layers are both unstable to the KHI and their evolutions are not symmetric about $y$-axis at $t=9.1$.
Subsequently, the plasma under the TS starts to stir significantly and the flows show different behaviors on both sides of $y$-axis.
At $t=9.46$, the previously $y$-symmetric TS front becomes oblique (see Figure \ref{fig2}(d2)).
Both shear layers are still unstable to the KHI (see Figure \ref{fig2}(e2)).
The left one starts to rotate after this moment, and, at $t=9.66$, it evolves into a horizontal layer that still satisfies the condition of the KHI (see Figures \ref{fig2}(c3)-(e3)).
Thereafter, similar behaviors of shear layers are observed and the oscillation amplitude of the CS bottom also grows.

We now examine the condition of the RTI in the loop-top region.
The RTI can be switched on if the perturbation wave along the density interface satisfies \citep{Hillier2016a,Carlyle2017,XieXiaoyan_2021_977}
\begin{equation}
\omega^2=-kg_\perp\frac{\rho_u-\rho_l}{\rho_u+\rho_l}+\frac{2k^2B^2_\parallel}{\rho_u+\rho_l}<0\,\label{eq:rti}
\end{equation} 
where $k$ and $\omega$ denote respectively the wave-number and frequency of the perturbation, $g_\perp$ is the acceleration of gravity perpendicular to the interface, $B_\parallel$ is the magnetic field strength parallel with the interface, and $\rho_u$ and $\rho_l$ are the upper and lower density, respectively. 
Equation \ref{eq:rti} means that the perturbation wave-length $\lambda=2\pi/k$ should be larger than the critical length $\lambda_c=4\pi B_\parallel^2/g_\perp\left(\rho_u-\rho_l\right)$ for the sake of initiating the RTI.
The density interface where the RTI might grow satisfies $\hat{{\bf e}}_y\cdot\nabla\rho>0$ to guarantee $\rho_u-\rho_l>0$ and its normal vector is $\hat{\bf n}_{\mathrm{rt}}=\nabla\rho/\left|\nabla\rho\right|$.
We can numerically calculate the local value of $\lambda_c$ after defining $B_\parallel^2\equiv B_x^2+B_y^2-\left({\bf B}\cdot\hat{\bf n}_{\mathrm{rt}}\right)^2$, $g_\perp\equiv\left|{\bf g}\cdot\hat{\bf n}_{\mathrm{rt}}\right|$, and $\rho_u-\rho_l\equiv\left|\nabla\rho\right|\Delta L$.
Because the perturbation wave-length should be smaller than the scale of the loop-top region, we locate all the positions satisfying $\lambda_c<0.1$ and find that they show a highly scattered distribution (see the white dots in Figure \ref{fig2}(a1)-(a3)). 
Moreover, moving with turbulent flows, these isolated positions vary dynamically in time (see the animation of Figure \ref{fig2}), which means that their lifetime might be too short to initiate the RTI.
In contrast, the main shear layers triggering the KHI typically extend $\sim 0.1L_0$ continuously in space. 
As the development of KHI, the shapes of the shear layers vary gradually in time but their spatial scales are approximately maintained (see Figure \ref{fig2}).
Meanwhile, in our previous simulations in which gravity is not included and the RTI effect is ignorable, the loop-top oscillations can still be observed \citep[see][]{WYL_2021_964}.
Therefore, we can conclude that the KHI dominates the symmetry breaking of the loop-top region and thus the initialization of CS oscillations. 

\subsection{Propagation properties of CS Oscillations\label{subsec:wavechr}}

\begin{figure*}[h]
\gridline{\fig{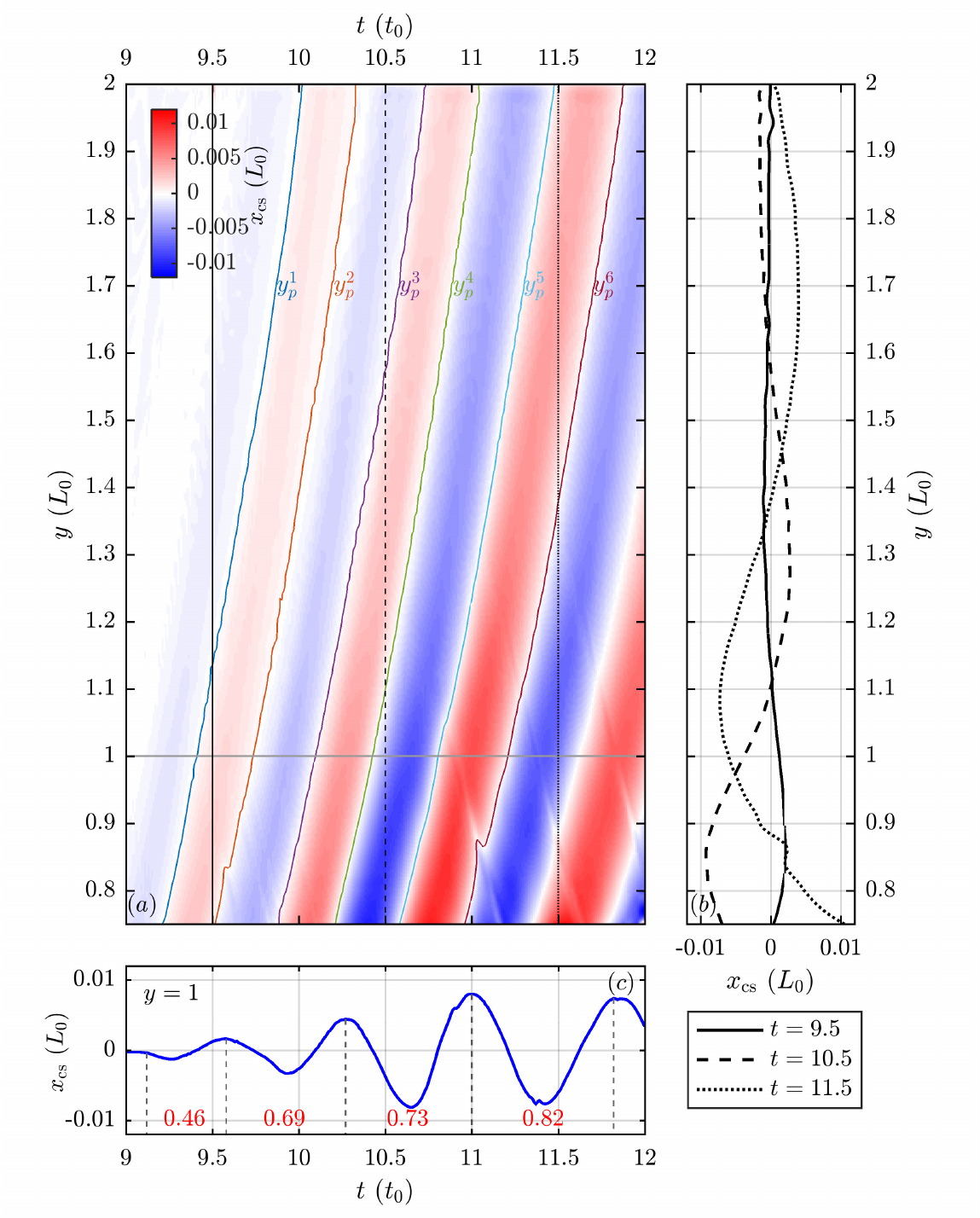}{0.7\textwidth}{}}
\caption{(a) The time-distance plot of the central positions of the CS $x_{\mathrm{cs}}\left(y\right)$ from $t=9$ to $12$.
(b) The profiles of the CS positions $x_{\mathrm{cs}}$ at $t=9.5$, $10.5$, and $11.5$.
(c) The temporal evolution of the CS position $x_{\mathrm{cs}}$ at $y=1$.
The red numbers in panel (c) record the periods of four oscillations.
The colored curves in Panel (a) are six contours of $x_{\mathrm{cs}}=0$.
The high-frequency noises in $x_\mathrm{cs}$ have been filtered.
An animation of this figure is available online. 
The video, with a duration of $19\,\mathrm{s}$, shows the dynamical evolution of $x_\mathrm{cs}\left(y\right)$.
\label{fig3}}
\end{figure*}

We further quantitatively analyze the propagation properties of the CS oscillation.
Based on the method introduced in Appendix \ref{sec:appendix}, we determine the central positions $x_\mathrm{cs}\left(y\right)$ of the CS (see Figures \ref{fig1}(b1)-(b6) and \ref{fig3}(b)).
The distance-time plot of the positions clearly shows some wave-like structures (Figure \ref{fig3}(a)).
During the first three oscillations, both the amplitude and the period grow (see Figure \ref{fig3}(c)), and the maximum oscillation amplitude reaches $\sim 0.01$.
Though damping as propagating upward, the oscillation can still be clearly detected at a high altitude (see Figure \ref{fig3}(a)).

\begin{figure*}[h]
\gridline{\fig{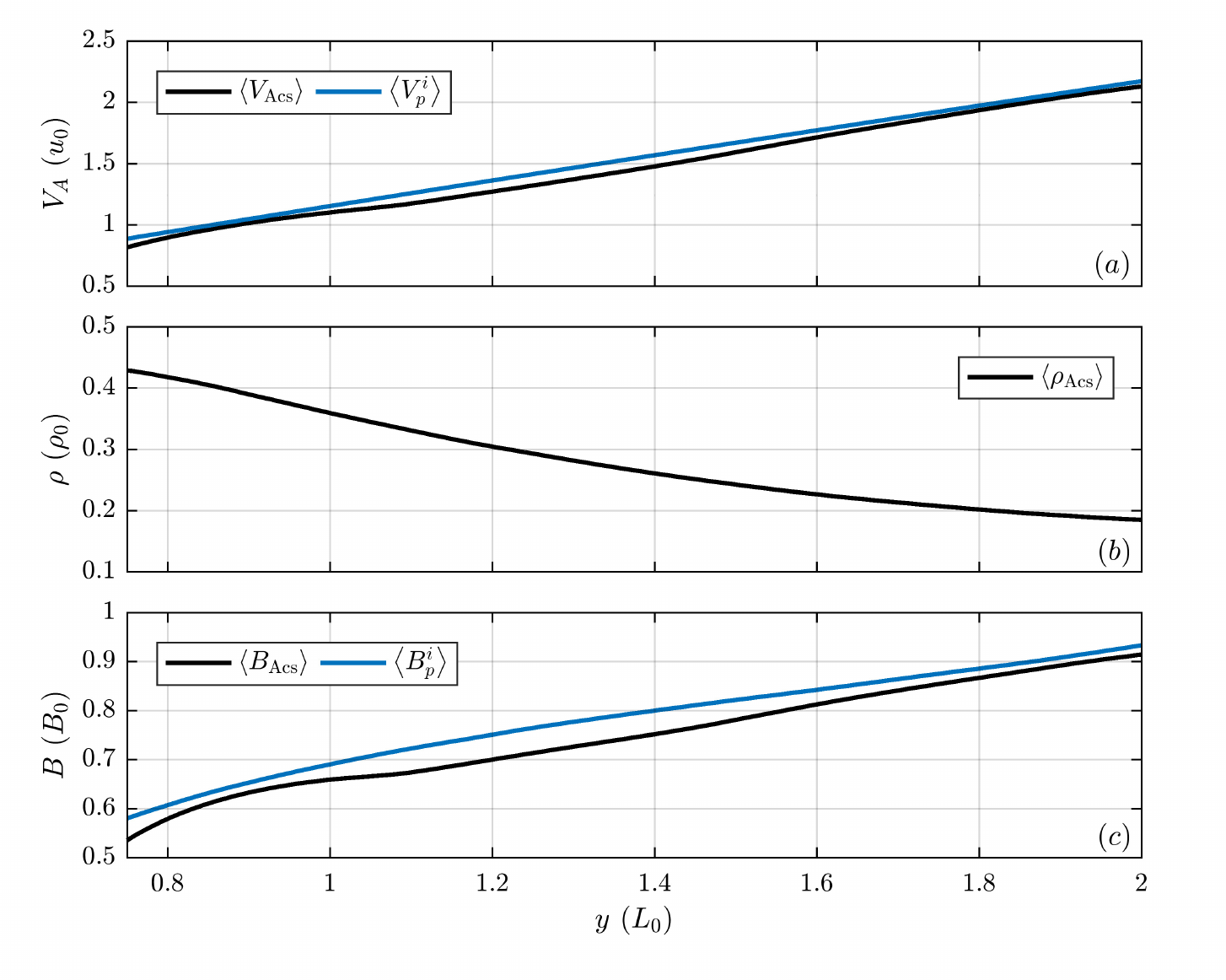}{0.6\textwidth}{}}
\caption{(a) The profiles of time-averaged local Alfv\'{e}n speed at the CS outer boundary (the black curve) and the averaged propagation speeds of the CS oscillation (the blue curve).
(b) The time-averaged density at the CS outer boundary.
(c) The distributions of time-averaged magnetic strength at the CS outer boundary (the black curve) and the magnetic field strength evaluated by $\left<B_p^i\right>=\left<V_p^i\right>\sqrt{\left<\rho_{\mathrm{Acs}}\right>}$ (the blue curve).
The CS outer boundaries are defined by $x=\pm 0.05$ and $y>0.75$ (see Figure \ref{fig1}(d6)) and  $V_{\mathrm{Acs}}$, $\rho_{\mathrm{Acs}}$, and $B_\mathrm{Acs}$ are averaged values on both boundaries.
$\left<\cdot\right>$ denotes taking time-average in duration $t\in\left[9,12\right]$.
\label{fig4}}
\end{figure*}

To estimate the propagation speed of the perturbations, we take six contours in Figure \ref{fig3}(a) and denote them by $y^i_p\left(t\right)$, where $i=1,2,\cdots,6$ (see Figure \ref{fig3}(a)).
These contour lines track the propagation of the perturbations causing a zero-displacement at the $x$-direction.
Note that, the propagation curves contain small noises caused by the motion of magnetic islands.
We fit the contour lines by the power function to clean the noises and then take their time derivatives to obtain the propagation speeds, namely, $V^i_p\left(y\right)$.
The speeds calculated from six different tracks are very similar and we take their averaged value $\left<V^i_p\right>$ to evaluate their mean feature (see Figure \ref{fig4}(a)). 
The speed increases from $0.89$ to $2.17$ with height, corresponding to the real values from $388$ to $946\,\mathrm{km/s}$.
Furthermore, we calculate the local Alfv\'{e}n speed at the CS outer boundary ($V_\mathrm{Acs}$).
It is found that the variations of $\left<V^i_p\right>$ with height are largely similar to the profile of the time-average Alfv\'{e}n speed $\left<V_\mathrm{Acs}\right>$ (see Figure \ref{fig4}(a)).

Considering the similarity between $\left<V_\mathrm{Acs}\right>$ and $\left<V^i_p\right>$, we propose a novel way to determine the magnetic field strength of the CS outer boundary.
Observationally, one can use the averaged propagation speeds of several CS oscillations, $\left<V^i_p\right>$, as a proxy of the local Alfv\'{e}n speed. 
With a combination of the plasma density estimated by differential emission measure (DEM), the magnetic field strength of the CS boundary is then derived as
\begin{equation}
\left<B_\mathrm{Acs}\right>=\left<V_\mathrm{Acs}\right>\sqrt{\left<\rho_\mathrm{Acs}\right>}\approx \left<V^i_p\right>\sqrt{\left<\rho_\mathrm{Acs}\right>}=\left<B^i_p\right>\,.\label{eq:Bcse} 
\end{equation}
Here, we verify this method using the numerical data (Figure \ref{fig4}).
The time-averaged value $\left<\rho_\mathrm{cse}\right>$ are used to simulate the density derived by the DEM analysis (Figure \ref{fig4}(b1) and (b2)).
We find that the profiles of $\left<B^i_p\right>$ we evaluated are basically in agreement with the distribution of $\left<B_\mathrm{Acs}\right>$ in height (see Figure \ref{fig4}(c)).

\section{Summary and Discussion\label{sec:sum}}

In this letter, we study the formation mechanism of the CS oscillations using MHD simulation.
The fully-developed reconnection presents a quasi-periodic feature.
Various downflow magnetic islands collide with the flare loop top and result in a turbulent cusp-shaped structure.
Following the collision of a relatively large island at the loop top, fewer and smaller islands are generated.
The KHI is then switched on in the interface of two sheared flows, from the persistent reconnection downflows.
The asymmetric flows, caused by the asymmetric KHI, finally drive large-amplitude swaying of the flare loop-top and thus the oscillations of the CS.

The CS oscillations, though damping with distance, are still detectable when they propagate to a higher height.
More importantly, the CS oscillations are proven to propagate with the local Alfv\'{e}n speed.
It thus can be used for evaluating the magnetic field strength of the outer boundary of the CS once we derive the local plasma density, for example, through the DEM technique. 
This method is well verified by our numerical data and is hopeful to be comparable with other methods \citep[e.g.,][]{Chen2011a,Nakariakov2001,LiuWei_2011_890,Tian2012,Zihao2020}.
However, note that this method only provides a zero-order approximation of magnetic field strength.
Observationally, the main challenge is how to accurately measure the propagation speed of the CS oscillations, which are largely limited by observational tempo-spatial resolution.
Our result shows that the magnetic field at the outer boundary of the CS increases with the height, which is different from real observations and recent simulations of the flux rope eruption that use decaying magnetic fields in height \citep[e.g.,][]{Chen2020a}.
The main reason is that we take advantage of a uniform background magnetic field along y-direction and the appearance of upward moving magnetic islands during the CS oscillations.

Differing from \cite{TakahashiTakuya_2017_824}, in which the oscillations above the flare loops were also reproduced but for the TS fronts, here, we mainly focus on the fundamental mechanism and properties of large-amplitude oscillations of the whole CS. 
The mechanism for the CS oscillations revealed here is different from that in \cite{XieXiaoyan_2021_977} who considered the effects of CME but did not include thermal conduction, radiation cooling, and background heating. 
They contributed the CS oscillations to the RTI at the bottom of CME.
Although excluding the flux rope eruption, our model provides much more fine structures at the loop top region.
We find that, the positions where the RTI may be initiated are spatially scattered and temporally varied and thus, compared with the RTI, the KHI plays a more important role in giving rise to the CS oscillations.

\acknowledgments
We would like to thank the anonymous referee for valuable suggestions.
This research is supported by the Natural Science Foundation of China grants 
11722325, 11733003, 11790303, and 11790300.

\appendix
\section{The CS structure\label{sec:appendix}}
\restartappendixnumbering

\begin{figure*}[h]
\gridline{\fig{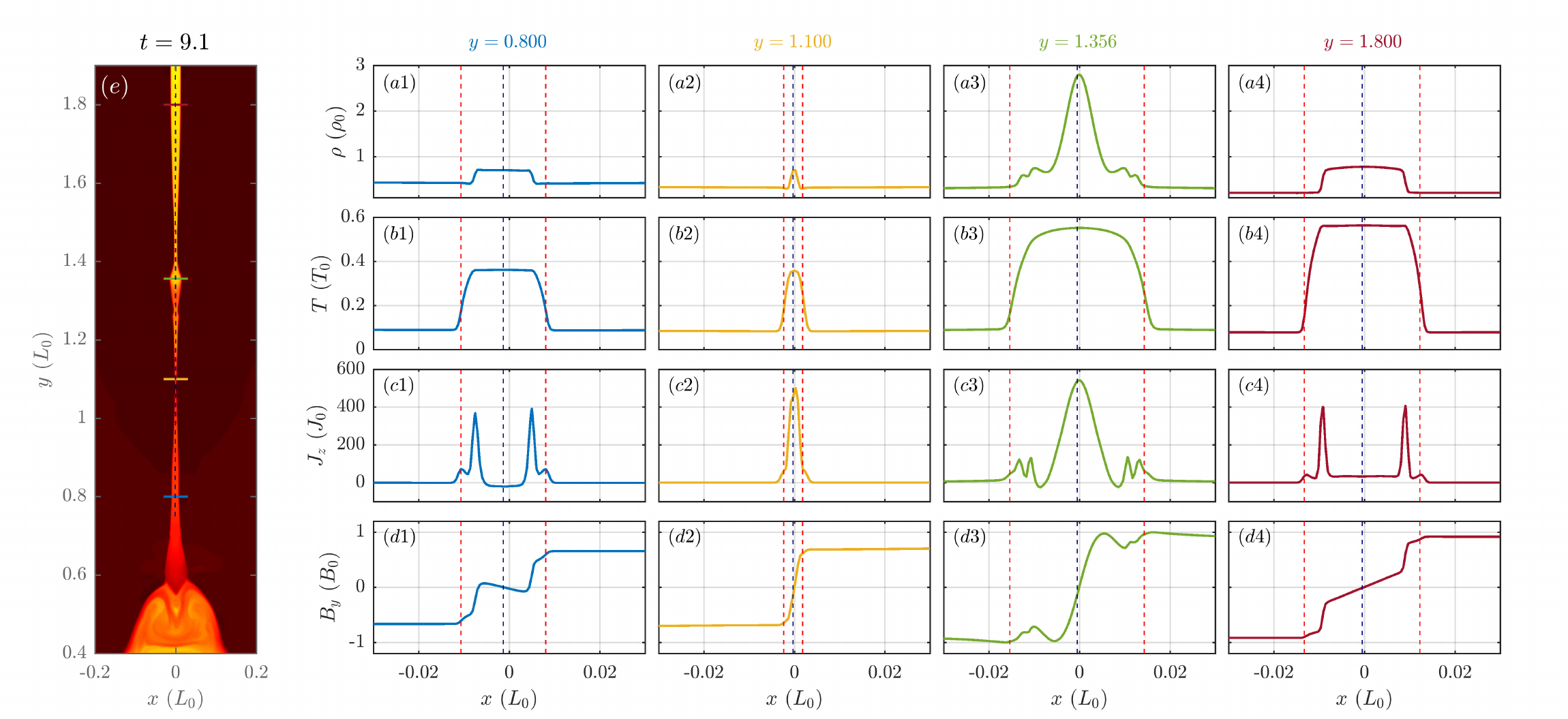}{1.0\textwidth}{}}
\caption{The CS structure at $t=9.1$. (e) The temperature distribution of the CS at $t=9.1$ (also see Figure \ref{fig1}(b1)).
The blue, yellow, green, and brown slits are taken at $y=0.8$, $1.1$, $1.356$, and $1.8$ across the CS, respectively.
(a1), (b1), (c1), and (d1) are respectively the profiles of $\rho$, $T$, $J_z$, and $B_y$ along the blue slit at $y=0.8$.
(a2)-(d2), (a3)-(d3), and (a4)-(d4) are the same as (a1)-(d1) but for different slits.
The red dashed lines mark the positions with the maximum temperature gradient ($\left|\partial T/\partial x\right|$) on both sides of the CS, while the blue dashed lines denote the average positions of the red ones. \label{figA1}}
\end{figure*}

We take four typical slits across the CS at $t=9.1$ to exhibits the detailed CS structures (see Figure \ref{figA1}(e)).
The red dashed lines mark the position with the maximum temperature gradients ($\left|\partial T/\partial x\right|$) on both sides of the CS, which approximately correspond to the discontinuity fronts of the slow-mode shocks \citep[also see][]{MeiZhiXing_2012_614}.
We take their average $x$-coordinate as the central position $x_\mathrm{cs}\left(y\right)$ of the CS (see the blue dashed lines in Figure \ref{figA1}).

Between the slow shocks, the polarity of magnetic field reverses rapidly in space (see Figures \ref{figA1}(d1)-(d4)). 
However, differing from standard CS models (e.g., the Harris-sheet or the Petschek-sheet), the CS structure shows complex details.
Near the principal X-point ($y=1.1$), the CS is thinnest and shows a Gauss-type distribution.
At $y=0.8$ and $y=1.8$, on both sides of the CS, the slow-shock fronts cause rapid decreases of magnetic field strength and thus induce two localized strong currents, between which a lower current plateau forms.
$J_z$ near $x=0$ can be either negative (Figure \ref{figA1}(c1), also see \cite{MeiZhiXing_2012_614}) or positive (Figure \ref{figA1}(c4)).
Near the center of a magnetic island ($y=1.356$), the density and current increase significantly (Figure \ref{figA1}(a3) and (c3)), while the temperature and magnetic field profiles are similar with other slits (Figure \ref{figA1}(b3) and (d3)).

For all four positions, however, one can see that the region of the CS is well enclosed by the red dashed lines and is approximately symmetric about the blue dashed lines.
Therefore, we determine the central position of the CS by the blue dashed line.

\bibliography{refs}{}
\bibliographystyle{aasjournal}

\end{document}